\documentclass[aps,prl,twocolumn,superscriptaddress,floatfix]{revtex4}
\usepackage{graphicx,graphics}
\usepackage{dcolumn}
\usepackage{amsmath,amssymb,amsfonts}
\usepackage{latexsym,verbatim}
\usepackage{bm}
\usepackage{color}
\usepackage{ulem}
\def\be{\begin{equation}}
\def\ee{\end{equation}}
\def\ber{\begin{eqnarray}}
\def\eer{\end{eqnarray}}

\begin{document}
\title{Drude weight, cyclotron resonance, and the Dicke model of graphene cavity QED}
\author{Luca Chirolli}
\affiliation{NEST, Scuola Normale Superiore and Istituto Nanoscienze-CNR, I-56126 Pisa, Italy}	
\author{Marco Polini}
\affiliation{NEST, Istituto Nanoscienze-CNR and Scuola Normale Superiore, I-56126 Pisa, Italy}
\author{Vittorio Giovannetti}
\affiliation{NEST, Scuola Normale Superiore and Istituto Nanoscienze-CNR, I-56126 Pisa, Italy}
\author{Allan H. MacDonald}
\affiliation{Department of Physics, University of Texas at Austin, Austin, Texas 78712, USA}

\begin{abstract}
The Dicke model of cavity quantum electrodynamics is approximately realized in condensed matter
when the cyclotron transition of a two-dimensional electron gas is nearly resonant with a
cavity photon mode.  We point out that in the strong coupling limit 
the Dicke model of cavity cyclotron resonance
must be supplemented by a term that is quadratic in the cavity photon field and 
suppresses the model's transition to a super-radiant state.  We develop the theory of graphene cavity
cyclotron resonance and show that the quadratic term, which is absent in graphene's 
low-energy Dirac model Hamiltonian, is in this case dynamically generated 
by virtual inter-band transitions.   
\end{abstract}
\maketitle

\noindent
{\it Introduction---}Distinct but degenerate dipole transitions between matter states can behave cooperatively
in an optical cavity because the photon continuum is discretized~\cite{walter_repprogphys_2006}.   
This physics is often captured approximately by the Dicke model~\cite{dicke_pr_1954}  which describes a non-dissipative closed system of identical two-level systems interacting with a 
single-mode radiation field.  For a sufficiently strong light-matter coupling constant, the thermodynamic 
limit of the Dicke model exhibits a second-order quantum phase transition to 
a super-radiant ground state (SPT)~\cite{SPT} with macroscopic photon occupation and coherent atomic polarization.
Wide interest in these SPTs has emerged recently in the context of 
{\it circuit} quantum electrodynamics (QED)~\cite{circuitQED}.  It has been shown in particular 
that superconducting flux qubits inductively coupled to a transmission line resonator display a SPT characterized by a doubly-degenerate ground state with macroscopic photon occupation~\cite{nataf_prl_2010}. There is
also interest in the possibility of a related phase transition for Cooper-pair boxes capacitively coupled to a transmission line resonator~\cite{nataf_naturecommun_2010,viehmann_prl_2011}.  Ultracold atom gases in optical cavities~\cite{brennecke_nature_2007,baumann_nature_2010}, 
in which spontaneous breaking of translational symmetry has been reported~\cite{baumann_nature_2010},
have provided still another interesting new application of Dicke model physics.  

When an external magnetic field is applied to a two-dimensional (2D) electron system, transitions 
between states in full and empty Landau levels (LLs) are dispersionless, mimicking atomic transitions and enabling a 
condensed matter realization of the Dicke model.
In particular, recent pioneering theoretical~\cite{hagenmuller_prb_2010} and experimental~\cite{scalari_science_2012} work 
has shown that these systems can be driven toward the ultrastrong coupling~\cite{ultrastrong} limit 
by tuning the cyclotron transition energy of an ordinary 
parabolic-band 2D electron gas to resonance with 
the photonic modes of a terahertz metamaterial.

The light-matter interaction in the Dicke Hamiltonian is linear in the vector potential ${\bm A}_{\rm em}$ of 
the cavity.  For condensed matter states described by parabolic band models,
a quadratic ${\bm A}^2_{\rm em}$ term whose strength is related to the system's 
Drude weight~\cite{Pines_and_Nozieres} and f-sum rule~\cite{Giuliani_and_Vignale}, 
also emerges naturally from minimal coupling.  It has long been
understood~\cite{nogotheorems} that the Dicke model's SPT is suppressed when the quadratic terms 
are retained.  Demonstrations of this property are often referred to as {\it no-go} theorems.
(Standard {\it no-go} theorems do not apply~\cite{nonequilibrium} to ultracold atoms which are driven by an external pump field and subject to significant cavity losses.)

We focus here on the case of cyclotron transitions in a graphene 2D electron system.
Electronic states near the neutrality point of a graphene sheet~\cite{review2007} 
exhibit linear crossing between conduction and valence bands. 
The effective low-energy theory for electrons in graphene has  
a 2D massless Dirac fermion (MDF) Hamiltonian~\cite{castroneto_rmp_2009}, which is
linear in momentum ${\bm p}$. 
As first noted in Ref.~\cite{hagenmuller_arxiv_2011}, minimal coupling applied to the MDF Hamiltonian
does not generate  a term proportional to ${\bm A}^2_{\rm em}$.
Cyclotron resonance in this material, which has been extensively investigated experimentally and theoretically
over the past decade~\cite{goerbig_rmp_2011}, seems therefore to 
provide an example of an active medium which could enable a SPT~\cite{hagenmuller_arxiv_2011}
when the graphene sheet is embedded in a cavity. 
Indeed, recent experimental advances have made it possible to monolithically integrate graphene with optical 
microcavities~\cite{engel_naturecommun_2012,furchi_nanolett_2012}, paving the way for cavity QED at the nanometer scale with graphene as an active medium.  In this work we demonstrate however that 
in the strong coupling regime the Dicke model for graphene cavity cyclotron resonance
must be supplemented by a quadratic term that is dynamically generated by inter-band 
transitions and again implies a {\it no-go} theorem.  These {\it no-go} theorems are 
ultimately a consequence of gauge invariance and apply whenever this symmetry is unbroken. 

\noindent
{\it Gauge invariance and SPTs---}We consider an electronic system in $D$ spatial dimensions 
coupled to an electromagnetic (e.m.) field with a single privileged mode described by a vector potential ${\bm A}({\bm r}, t)$.  
We argue below that the {\it no-go} theorem
for the SPT requires only unbroken gauge symmetry.  
Our ideas are most clearly spelled out when ${\bm A}$ is treated classically. Quantization of the e.m. field can be easily carried out in the final step. Light-matter interactions are
described by minimal coupling: ${\bm p}_i \to {\bm p}_i +e {\bm A}({\bm r}_i,t)/c$, where ${\bm p}_i$ 
is the canonical momentum of the $i$-th electron and $-e$ is the electron charge. 
The Hamiltonian of a light-matter system can always be written as ${\hat {\cal H}}[{\bm A}] = {\hat {\cal H}}_{\rm mat}[{\bm A}] + {\cal H}_{\rm em}[{\bm A}]$, 
where ${\hat {\cal H}}_{\rm mat}[{\bm A}]$ contains all the electronic degrees of freedom treated quantum mechanically, while ${\cal H}_{\rm em}[{\bm A}]$ is the classical energy density of the e.m. field.
The spontaneous coherent photon state is the ground state when the total energy 
is lowered by introducing a finite static vector potential.   
Since ${\cal H}_{\rm em}[{\bm A}]$ is a positive-definite quadratic form of ${\bm A}$, the
instability can occur only if the second derivative of the matter energy with 
respect to ${\bm A}$ is negative for static ${\bm A}$.   

We therefore consider the variation of the matter energy $\Delta E_{\rm mat} \equiv E_{\rm mat}[\delta {\bm A}] - E_{\rm mat}[{\bm 0}]$ due to an infinitesimal variation of the static vector potential:
\be\label{eq:deltaE}
\Delta E_{\rm mat} = \int d^D{\bm r}~\delta{\bm A} \cdot \big\langle \delta {\hat {\cal H}}_{\rm mat}[{\bm A}]/\delta {\bm A} \big\rangle~.
\ee
The quantity $\delta {\hat {\cal H}}_{\rm mat}[{\bm A}]/\delta {\bm A}$ is~\cite{currentcaveat} the physical current operator, ${\hat {\bm J}}_{\rm phys}({\bm r})$.  It is convenient in perturbative analyses
 to split the current into paramagnetic and diamagnetic contributions, which can be defined by the expansion of ${\hat {\bm J}}_{\rm phys}({\bm r})$ in powers of $\delta {\bm A}$. 
For the $\mu$-th Cartesian component of the current operator this expansion reads
\be\label{eq:currentoperator}
{\hat J}^{(\mu)}_{\rm phys}({\bm r}) = {\hat J}^{(\mu)}_{\rm p}({\bm r}) + \frac{e}{c}\int d^D{\bm r}'{\hat K}^{\mu \nu}({\bm r}, {\bm r}')\delta A_\nu({\bm r}')~,
\ee
where ${\hat J}^{(\mu)}_{\rm p}({\bm r}) = \delta{\hat {\cal H}}_{\rm mat}[{\bm A}]/\delta A_\mu({\bm r}) |_{{\bm A} ={\bm 0}}$ 
is the paramagnetic current-density operator, 
${\hat K}^{\mu \nu}({\bm r}, {\bm r}') = (c/e)\delta^2 {\hat {\cal H}}_{\rm mat}[{\bm A}]/\delta A_\mu({\bm r})\delta A_\nu({\bm r}')|_{{\bm A} ={\bm 0}}$, and the sum over repeated Greek indices is intended. 
The second term on the r.h.s. of Eq.~(\ref{eq:currentoperator}) is the diamagnetic
contribution to the current-density operator.

 We now evaluate the expectation value of ${\hat J}^{(\mu)}_{\rm phys}({\bm r})$ to first order in $\delta {\bm A}$: i)
the expectation value of the paramagnetic current-density operator, $ J^\mu_{\rm p}({\bm r})$, can be calculated using
linear response theory~\cite{Pines_and_Nozieres,Giuliani_and_Vignale}:
we find $ J^\mu_{\rm p}({\bm r}) = V^{-1}\sum_{\bm q} J^\mu_{\rm p}({\bm q})e^{i {\bm q}\cdot {\bm r}} +{\rm c.c.}$ with $V = L^D$  the $D$-dimensional volume and
$J^\mu_{\rm p}({\bm q}) = (e/c)\chi^{\mu\nu}(q)\delta A_\nu({\bm q})$, where the tensor $\chi^{\mu\nu}(q)$ is the {\it static} paramagnetic current-current response function. 
This current corresponds to the linear term in energy that is retained in the Dicke model.  
The diamagnetic current operator in Eq.~(\ref{eq:currentoperator}) corresponds to the quadratic 
term neglected in the Dicke model.  It already contains 
one power of $\delta {\bm A}$. Its contribution to the linear response current can therefore be obtained by 
evaluating its expectation value, 
$K^{\mu\nu}(|{\bm r}- {\bm r}'|) \equiv \langle {\hat K}^{\mu \nu}({\bm r}, {\bm r}')\rangle$, where $\langle \dots \rangle$ refers now to the ground state of ${\hat {\cal H}}_{\rm mat}$ at $\delta {\bm A} ={\bm 0}$~\cite{crystalwarnings}.  

Using these results, we find that $\langle {\hat J}^{(\mu)}_{\rm phys}({\bm r})\rangle = (e/c) V^{-1}\sum_{\bm q} \Xi^{\mu\nu}(q)\delta A_\nu({\bm q})e^{i {\bm q}\cdot {\bm r}} +{\rm c.c.}$, where we have introduced $\Xi^{\mu\nu}(q) \equiv \chi^{\mu\nu}(q) + K^{\mu\nu}(q)$, $K^{\mu\nu}(q)$ being the Fourier transform of $K^{\mu\nu}(|{\bm r}- {\bm r}'|)$.
The corresponding change in matter energy up to second order in $\delta {\bm A}$ is
$\Delta E_{\rm mat} \propto V^{-1}\sum_{\bm q}\Xi^{\mu\nu}(q)\delta A_\mu({\bm q}) \delta A_\nu(-{\bm q})$. The key point now is to realize~\cite{Pines_and_Nozieres,Giuliani_and_Vignale} that, if gauge invariance is unbroken, there cannot be a current response to a static uniform potential~\cite{moredetails}.  The diamagnetic contribution to the current-current response is crucial to cancel the paramagnetic contribution and can {\it never} be neglected:
\be\label{nogotheorem}
\lim_{q \to 0} K^{\mu\nu}(q) = - \lim_{q \to 0}\chi^{\mu\nu}(q)~.
\ee
This result, which is the most important of this Section, implies that, in the absence of broken gauge invariance, 
the change in matter energy due to a static and quasi-homogeneous vector potential must vanish~\cite{examples}. 
SPTs therefore cannot occur unless gauge invariance is broken, as occurs for example in the case of superconductors.

\begin{figure}[t]
\begin{center}
\includegraphics[width=0.70\linewidth]{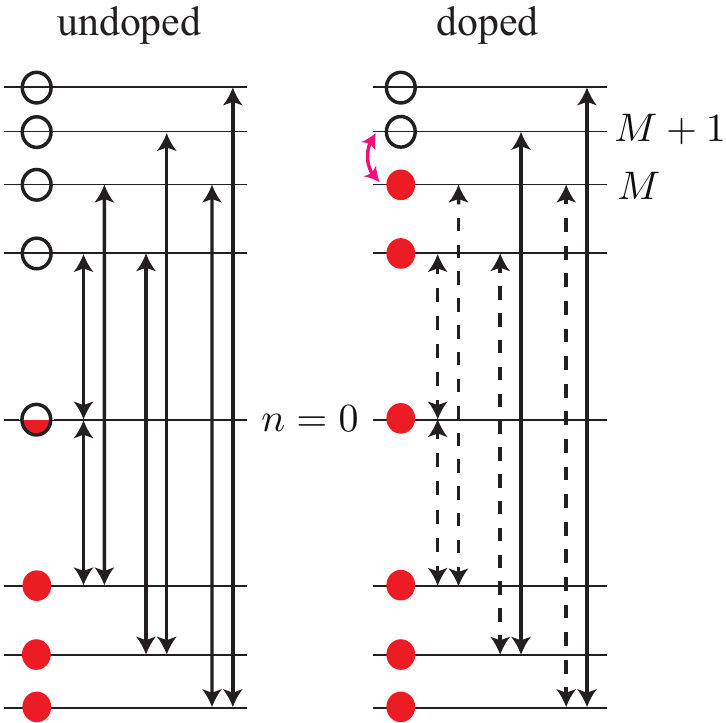}
\caption{(color online) Left panel: Dipole allowed transitions in a neutral graphene sheet in the presence of 
an external magnetic field. The horizontal lines denote the unevenly-spaced Landau levels of massless Dirac fermions.
The filled valence band levels (band label $\lambda=-1$) and the empty conduction band levels
(band label $\lambda=+1$) are 
indicated by filled (empty) circles.  The zero-energy ($n=0$) Landau level is formed 
partly from the valence band and partly from the conduction band and is half filled
in a neutral system. Right panel: In a doped graphene sheet with a Fermi level $\varepsilon_{\rm F}$ placed between 
the conduction band $n=M$ and the $n = M+1$ Landau levels, there is an allowed cyclotron 
transition within the conduction band (magenta line) while some of the lower energy inter-band transitions, indicated by dashed lines, 
are Pauli-blocked.  There is a clear energetic separation between the lowest energy (unblocked) intra-band transition 
and the (unblocked) inter-band transitions.\label{fig:cyclotronresonance}}
\label{fig1}
\end{center}
\end{figure}

\noindent
{\it Cavity cyclotron resonance in graphene---}We now focus on a 2D electron gas in a graphene sheet exposed
to a strong quantizing magnetic field ${\bm B} =B {\hat {\bm z}}$~\cite{eeinteractions}.  When 
Zeeman coupling is neglected, the LLs of graphene have two-fold spin and valley 
degeneracies and therefore overall degeneracy $N_{\rm f}=4$~\cite{goerbig_rmp_2011}.
The single-spin single-valley MDF Hamiltonian reads 
${\cal H}_0= v_{\rm D} {\bm \sigma} \cdot {\bm \Pi}({\bm r})$, where $v_{\rm D} \approx 10^6~{\rm m}/{\rm s}$ is the Dirac velocity, ${\bm \sigma} = (\sigma^x,\sigma^y)$ is a 2D vector of Pauli matrices, 
${\bm \Pi}({\bm r}) = -i\hbar \nabla_{\bm r} + e{\bm A}_0({\bm r})/c$ is the gauge-invariant dynamical momentum, and ${\bm A}_0({\bm r})$ is the static vector potential that describes the quantizing magnetic field. We work in the Landau gauge ${\bm A}_0({\bm r})  = (-By,0,0)$. A complete set of eigenfunctions is provided by the two-component pseudospinors~\cite{goerbig_rmp_2011}
\begin{equation}\label{eq:pseudospinors}
\langle {\bm r}|\lambda,n,k\rangle = \left(
\begin{array}{c}
\psi^{(A)}_{\lambda,n,k}({\bm r}) \vspace{0.1 cm}\\ \psi^{(B)}_{\lambda,n,k}({\bm r})
\end{array}
\right) =  
\frac{e^{ikx}}{\sqrt{L}} 
\left(
\begin{array}{c}
C^-_n\varphi_{n-1,k}(y) \vspace{0.1 cm}\\ \lambda C^+_n\varphi_{n,k}(y)
\end{array}
\right)~,
\end{equation}
where $\lambda= +1$ ($-1$) for conduction (valence) band levels, $n \in {\mathbb N}$
is the intra-band  LL index, and $k$ is the eigenvalue of the magnetic translation operator in the ${\hat {\bm x}}$ direction. 
In Eq.~(\ref{eq:pseudospinors}) $C_n^{\pm}=\sqrt{(1\pm \delta_{n,0})/2}$ and 
$\varphi_{n,k}(y) $ are the normalized eigenfunctions of the parabolic-band Landau problem~\cite{Giuliani_and_Vignale}. 
The form of the coefficients $C_n^{\pm}$ guarantees that the pseudospinor corresponding to the $n=0$ LL has weight 
only on one sublattice. Each LL has a degeneracy ${\cal N}= N_{\rm f} S/(2\pi \ell_B^2)$, where $S=L^2$ is the sample area.
The pseudospinor (\ref{eq:pseudospinors}) is an eigenstate of the Hamiltonian ${\cal H}_0$ with eigenvalue~\cite{goerbig_rmp_2011} $\varepsilon^{(0)}_{\lambda, n} = \lambda \hbar \omega_{\rm c} \sqrt{n}$, where we have introduced the MDF cyclotron frequency $\omega_{\rm c}= v_{\rm D}\sqrt{2}/\ell_B$ and the magnetic length $\ell_B =[\hbar c/(eB)]^{1/2}$. 

The Hamiltonian which describes coupling between MDFs and light in the cavity 
does not contain a quadratic term: 
${\cal H}_{\rm int} = v_{\rm D} (e/c) \, {\bm \sigma} \cdot {\bm A}_{\rm em}({\bm r})$. 
For future purposes we introduce the notations $\sigma^\pm = (\sigma^x \pm i \sigma^y)/2$ and 
$A^{\pm}_{\rm em}({\bm r}) = A^x_{\rm em}({\bm r}) \pm i A^y_{\rm em}({\bm r})$. In what follows we neglect~\cite{hagenmuller_arxiv_2011} the spatial variation of the e.m. field in the cavity, $A^{\pm}_{\rm em}({\bm r}) \to A^{\pm}_{\rm em}$, since the photon wavelength is normally much larger than other length scales in the problem.
In this quasi-uniform approximation we can easily evaluate the matrix elements of ${\cal H}_{\rm int}$ between the unperturbed pseudospinors $(\ref{eq:pseudospinors})$:
\begin{eqnarray}\label{matrixelements}
\langle \lambda',n',k'| {\cal H}_{\rm int}|\lambda,n,k\rangle &=& 
\frac{e v_{\rm D}}{c}\delta_{k,k'}\big(\lambda C^-_{n'}C^+_n\delta_{n', n+1}A^{-}_{\rm em}\nonumber\\
&+&\lambda' C^+_{n'}C^-_n \delta_{n', n-1}A^{+}_{\rm em}\big)~.
\end{eqnarray}

The strong coupling limit is most easily obtained when the Fermi energy $\varepsilon_{\rm F}$ lies within one of the bands;
we consider the case in which it lies in the  
conduction band ($\lambda = +1$) between the LL with index $n= M$, which is fully occupied, and 
the LL with index $n = M +1$, which is at least partially empty.
(See Fig.~\ref{fig:cyclotronresonance}.) The Dicke model of cavity cyclotron resonance includes 
only the intra-band $n=M$ to $n=M+1$ transition
and acts in the $2^{\cal N}$-fold subspace spanned 
by $\{|+, M, k\rangle, |+, M+1, k\rangle, k =1 \dots {\cal N}\}$, neglecting
inter-band transitions~\cite{hagenmuller_arxiv_2011}. 
Using Eq.~(\ref{matrixelements}) and introducing a
set of Pauli matrices $\{\openone_k, \tau^z_k, \tau^\pm_k, k = 1 \dots {\cal N}\}$ which act 
in this two-level-system subspace leads to the following pseudospin Hamiltonian:
%
\begin{equation}\label{onebandmodel}
{\cal H}_{\rm eff} = \sum_{k=1}^{\cal N} \left({\cal E}_M\openone_k - \frac{\Omega_M}{2}\tau^z_k +\epsilon_{\rm em}\tau^+_k+\epsilon^*_{\rm em}\tau^-_k\right)~,
\end{equation}
where ${\cal E}_M = \hbar\omega_{\rm c}(\sqrt{M}+\sqrt{M+1})/2$, 
$\Omega_M = \hbar\omega_{\rm c}(\sqrt{M+1}-\sqrt{M})$, and $\epsilon_{\rm em} =e v_{\rm D} A^-_{\rm em}/(2 c)$. 
In this model the occupied conduction-band LL shifts down in energy by
\begin{equation}\label{spuriousshift}
\Delta E^{({\rm intra})}_M = - {\cal N}\left(\frac{e v_{\rm D}}{2 c}\right)^2 \frac{{\bm A}^2_{\rm em}}{\hbar \omega_{\rm c}}(\sqrt{M+1} +\sqrt{M})~
\end{equation}
in the limit of a static vector potential, in violation of gauge invariance as explained in the previous section.
This is the origin of the SPT found in Ref.~\cite{hagenmuller_arxiv_2011}. 
The correct effective matter Hamiltonian ${\cal H}_{\rm mat}$ for graphene cavity cyclotron
resonance must repair this defect.

The Dicke model misses a diamagnetic contribution to ${\cal H}_{\rm mat}$, which, according to (\ref{nogotheorem}), 
must precisely cancel the spurious energy shift (\ref{spuriousshift}). 
To derive this term we first recognize the intrinsic two-band nature of graphene (see Fig.~\ref{fig:cyclotronresonance}). 
A generic valence band state $|-,n,k\rangle$
is coupled by the e.m. field to two states in conduction band: $|+,n+1,k\rangle$ and $|+,n-1,k\rangle$. 
We first consider the undoped limit in which 
all valence band states $|-,n,k\rangle$ are occupied.  Because the Dirac model applies over a large but 
finite energy region we must apply a cut-off by occupying 
valence band levels with $0 \leq n \leq \nu_{\rm max}$. 
Treating the e.m. field again by second-order perturbation theory, 
we find the following change in matter energy for an undoped graphene sheet in a quantizing magnetic field:
\begin{eqnarray}\label{2ndorderPTvalenceband}
\Delta E_{\rm undoped} &=&\sum_{k=1}^{\cal N}\sum_{n=0}^{\nu_{\rm max}}\left[p_{n}\frac{|\langle +,n+1,k|{\cal H}_{\rm int}|-,n,k\rangle|^2}{\varepsilon^{(0)}_{-,n} - \varepsilon^{(0)}_{+,n+1}} \right.\nonumber\\
&+&\left.p_{n-1}\frac{|\langle +,n-1,k|{\cal H}_{\rm int}|-,n,k\rangle|^2}{\varepsilon^{(0)}_{-,n} - \varepsilon^{(0)}_{+,n-1}}\right]~,
\end{eqnarray}
where $p_n = 1 -\delta_{n,0}/2$. (The factor $p_n$ takes care of 
transitions involving the $n=0$ LL, which is half filled.)
Using Eq.~(\ref{matrixelements}) for the matrix elements we can write Eq.~(\ref{2ndorderPTvalenceband}) more explicitly:
\begin{equation}\label{largecutoffcontribution}
\Delta E_{\rm undoped} = -{\cal N} \left(\frac{ev_{\rm D}}{2c}\right)^2\frac{{\bm A}^2_{\rm em}}{\hbar\omega_{\rm c}}[1 + F(\nu_{\rm max})]~,
\end{equation}
with $F(\nu) \equiv \sum_{n=1}^{\nu}\left(\frac{1}{\sqrt{n}+\sqrt{n+1}}+\frac{1}{\sqrt{n}+\sqrt{n-1}}\right)$. 
Once again, this large negative contribution to the change in matter energy is spurious.  It is present 
because the Dirac model with a rigid ultraviolet cut-off $\nu_{\rm max}$ breaks gauge invariance~\cite{abedinpour_prb_2011}.  
When a model that is correct at atomic length scales, for example a $\pi$-band tight-binding model, is used instead, a static vector potential merely reassigns momentum labels within the full valence band.  
We compensate exactly for this deficiency of the Dirac model at its ultraviolet cut-off scale by adding the positive 
quantity $-\Delta E_{\rm undoped}$ to the change in matter energy. 

We now reconsider the situation analyzed earlier in which the Fermi energy $\varepsilon_{\rm F}$ lies in conduction band ($\lambda = +1$) between LLs with indices $n= M$ and $n = M +1$, but account for inter-band transitions.  
The {\it inter-band} correction to the energy shift 
$\Delta E^{({\rm intra})}_M$ in Eq.~(\ref{spuriousshift}), 
can be calculated using an expression which is equivalent to Eq.~(\ref{2ndorderPTvalenceband}) 
but accounts for {\it Pauli blocking} of transitions to occupied conduction-band states. 
The final result for the inter-band contribution is given by
\begin{eqnarray}
\Delta E^{({\rm inter})}_M  &=& - \Delta E_{\rm undoped} -{\cal N} \left(\frac{ev_{\rm D}}{2c}\right)^2\frac{{\bm A}^2_{\rm em}}{\hbar\omega_{\rm c}}\nonumber\\
&\times&[F(\nu_{\rm max}) - F(M)]~, \nonumber \\
&=& {\cal N} \left(\frac{ev_{\rm D}}{2c}\right)^2\frac{{\bm A}^2_{\rm em}}{\hbar\omega_{\rm c}}[1 + F(M)]~,
\end{eqnarray}
where the term $- \Delta E_{\rm undoped}$ takes care of the Dirac model regularization and Eq.~(\ref{largecutoffcontribution}) has been used in the last equality. After noticing that $1+F(M) = \sqrt{M+1} + \sqrt{M}$, we finally obtain
\begin{equation}\label{interbandcontributionfinal}
\Delta E^{({\rm inter})}_M  = {\cal N} \left(\frac{ev_{\rm D}}{2c}\right)^2\frac{{\bm A}^2_{\rm em}}{\hbar\omega_{\rm c}}(\sqrt{M+1} +\sqrt{M})~.
\end{equation}
The quantity $\Delta E^{({\rm inter})}_M$ is a dynamically-generated inter-band diamagnetic contribution to the effective Hamiltonian ${\cal H}_{\rm eff}$, which: i) is independent of cut-off $\nu_{\rm max}$ and ii) satisfies $\Delta E_{\rm mat} = \Delta E^{({\rm intra})}_M + \Delta E^{({\rm inter})}_M =0$, {\it i.e.} in the limit of a static vector potential, it precisely cancels the spurious shift (\ref{spuriousshift}) responsible for the Dicke model SPT.  

Because intra-band transition energies are much lower than inter-band transitions in the 
weak-field limit relevant to the strong coupling limit of cavity cyclotron resonance, we can neglect the 
frequency dependence of the dynamically generated quadratic term.  
This energy must be added to the effective matter Hamiltonian (\ref{onebandmodel}) for 
graphene cavity cyclotron resonance:
\begin{equation}\label{effectiveHamiltonian}
{\cal H}_{\rm eff} \to {\cal H}_{\rm eff} + S\frac{{\cal D}_M}{2\pi c^2} {\bm A}^2_{\rm em}~,
\end{equation}
where ${\cal D}_M = 4{\cal E}_M\sigma_{\rm uni}/\hbar$ is the Drude weight~\cite{examples} expressed in terms of the 
function  ${\cal E}_M$  introduced right after Eq.~(\ref{onebandmodel}).
This Hamiltonian 
is the starting point of the cavity QED theory of graphene cyclotron resonance.
The ${\bm A}^2_{\rm em}$ quadratic supplement to the Dicke model is always critical in the strong coupling limit.
Eq.~(\ref{effectiveHamiltonian}) is the most important result of this work.

\noindent
{\it Quantum Theory---}We can quantize the e.m. field by promoting the positive and negative Fourier amplitudes of ${\bm A}_{\rm em}$ to photon annihilation $a$ and creation $a^{\dag}$ operators: 
${\bm A}_{\rm em}={\cal A}{\bm \epsilon}(a+a^{\dag})$, where ${\bm \epsilon}$ is unit vector describing the polarization of the e.m. field and ${\cal A}=\sqrt{2\pi \hbar c^2/(\varepsilon\omega V)}$ with $V = S L_z$ the volume of the cavity ($L_z \ll L$ is the height of the cavity in the direction perpendicular to graphene) and $\varepsilon$ its dielectric constant.   
When a cavity model with frequency $\omega$ is 
nearly resonant with the cyclotron transition frequency $\Omega_M$, 
the total Hamiltonian (\ref{effectiveHamiltonian}) yields a Dicke model supplemented by an ${\bm A}^2_{\rm em}$ term:
\begin{eqnarray}\label{Dicke}
{\cal H}_{\rm Dicke}&=&\hbar\omega a^{\dag}a - \frac{\Omega_M}{2}\sum_{k=1}^{\cal N}\tau^z_k+\frac{g}{\sqrt{\cal N}}\sum_{k=1}^{\cal N}\tau^x_k(a+a^{\dag})\nonumber\\
&+&\kappa (a+a^{\dag})^2~,
\end{eqnarray}
where $g = \hbar\omega_{\rm c}\sqrt{2\sigma_{\rm uni}/(\varepsilon \omega L_z)}$ and $\kappa = \hbar {\cal D}_M/(\varepsilon \omega L_z)$. In writing Eq.~(\ref{Dicke}) we have assumed a specific polarization of the e.m. field, {\it i.e.} ${\bm \epsilon} = {\hat {\bm x}}$. In the thermodynamic ${\cal N} \to \infty$ limit the model (\ref{Dicke}) undergoes a SPT if the condition $\omega\Omega_M
(1+ 4\kappa/\omega)/(4g^2)<1$ is satisfied~\cite{nogotheorems,viehmann_prl_2011}. In our case, however, a SPT is forbidden because the following identity holds true:
\begin{equation}\label{gaugeinvariance}
g^2 = \kappa\Omega_M~.
\end{equation}
Equation~(\ref{gaugeinvariance}) specifically establishes
a {\it no-go} theorem for the occurrence of a SPT in the graphene cyclotron resonance cavity QED.
It is not a coincidence, and instead follows directly from 
the cancellation between paramagnetic and diamagnetic currents discussed in the first part of this work.
The paramagnetic response of the Hamiltonian (\ref{Dicke}) to a static and quasi-homogeneous e.m. field is, indeed, $(g^2/{\cal N})\lim_{\omega\to 0}\langle\langle \tau^x_{\rm tot}; \tau^x_{\rm tot}\rangle\rangle_\omega = 
- 2g^2/\Omega_M$, where $\tau^x_{\rm tot} = \sum_k\tau^x_k$ and $\langle\langle \tau^x_{\rm tot}; \tau^x_{\rm tot}\rangle\rangle_\omega = 2{\cal N}\Omega_M/(\omega^2-\Omega^2_M)$~\cite{abedinpour_prl_2007}. According to Eq.~(\ref{nogotheorem}), this paramagnetic contribution must be equal in magnitude and opposite in sign to the diamagnetic response of (\ref{Dicke}), which is simply $2\kappa$: {\it i.e.} it must satisfy $\kappa =  g^2/\Omega_M$, which coincides with Eq.~(\ref{gaugeinvariance}).

\noindent
{\it Summary---}We have derived a microscopic effective Hamiltonian - Eq.~(\ref{effectiveHamiltonian}) - 
for graphene cavity cyclotron resonance, highlighting in particular the role of gauge invariance and Drude weight. 
We find that the model must include a quadratic term which eliminates the possibility of super-radiant 
quantum phase transitions.  
The peculiar spectrum of the massless Dirac fermion Hamiltonian in a quantizing magnetic field, 
in which the Landau level spacing decreases with increasing energy, 
nevertheless makes this material  
particularly attractive for ultrastrong coupling between cavity photon modes and cyclotron transitions.
Electrons in graphene are thus an extremely intriguing active medium, 
and might pave the way for the realization of a gate-tunable generalized Dicke model.  
Light-matter interactions can be further enhanced by combining cavity-integrated graphene sheets with 
graphene-based plasmonic elements~\cite{koppens_nanolett_2011}. 
Our work does not touch upon the interesting possibility of realizing super-radiant quantum phases
when massless Dirac fermions are driven far away from equilibrium.

\noindent
{\it Acknowledgements---}We thank  Rosario Fazio for fruitful  discussions. 
Work in Pisa was supported by MIUR through the programs 
``FIRB IDEAS" - Project ESQUI (Grant No. RBID08B3FM) and ``FIRB - Futuro in Ricerca 2010" - Project PLASMOGRAPH (Grant No. RBFR10M5BT).   AHM was supported by Welch Foundation grant TBF1473
and by DOE Division of Materials Sciences and Engineering grant DE-FG03-02ER45958.

\begin{thebibliography}{77}
\bibitem{ultrastrong}
	In the ultrastrong limit the strength of the radiation-induced coupling between two consecutive LLs is of the same order as the bare cyclotron transition energy.
\bibitem{moredetails}
	Note that in a homogeneous and isotropic liquid $\Xi^{\mu\nu}(q)$ can be decomposed into a longitudinal 
	and a transverse part: $\Xi^{\mu\nu}(q) = (q_\mu q_\nu/q^2)\Xi_{\rm L}(q) + (\delta_{\mu\nu} -q_\mu q_\nu/q^2)\Xi_{\rm T}(q)$.
	The following properties hold true~\cite{Pines_and_Nozieres,Giuliani_and_Vignale}: $\Xi_{\rm L}(q) = 0$ for every $q$ 
	and $\lim_{q\to 0} \Xi_{\rm T}(q) =0$ (diamagnetic sum rule).
\bibitem{eeinteractions}
	Although it is clear that the gauge invariance condition (\ref{nogotheorem}) is valid in the presence of electron-electron interactions, 
	for simplicity in this Section we restrict the analysis to a non-interacting gas of MDFs. 
	At equilibrium, the impact of electron-electron interactions on the graphene cavity cyclotron resonance theory is merely 
	quantitative.
\end{thebibliography}
\end{document}